\documentclass{PoS}
\usepackage{graphicx}
\usepackage{dcolumn}
\usepackage{bm}
\usepackage{amssymb}
\usepackage{amsmath}
\usepackage{epsfig}    
\usepackage{color}
\usepackage{slashed}
\usepackage{makecell}
\usepackage{float}
\title{Theory and Phenomenology of Composite 2-Higgs Doublet Models}

\ShortTitle{C2HDM searches}

\author{{Stefania De Curtis}\\
        INFN, Sezione di Firenze, and Department of Physics and Astronomy, University of Florence, Via
G. Sansone 1, 50019 Sesto Fiorentino, Italy\\
        E-mail: \email{decurtis@fi.infn.it}}
        
 \author{{Stefano Moretti}\\
        School of Physics and Astronomy, University of Southampton, Southampton, SO17 1BJ, United Kingdom\\
        E-mail: \email{S.Moretti@soton.ac.uk}}       

\author{{Kei Yagyu}\\
        INFN, Sezione di Firenze, and Department of Physics and Astronomy, University of Florence, Via
G. Sansone 1, 50019 Sesto Fiorentino, Italy\\
        E-mail: \email{yagyu@fi.infn.it}}

 \author{\speaker{Emine Yildirim}\\
        School of Physics and Astronomy, University of Southampton, Southampton, SO17 1BJ, United Kingdom\\
        E-mail: \email{ey1g13@soton.ac.uk}}  

\abstract{In this talk, we report unitarity constraints and phenomenological studies at the Large Hadron Collider for the extra Higgs bosons of a Composite 2-Higgs Doublet Model.}

\FullConference{Prospects for Charged Higgs Discovery at Colliders\\
		3-6 October 2016\\
		Uppsala, Sweden}

\begin{document}

\section{Introduction}
After the discovery of a Higgs like state at the mass of 125 GeV  at the Large Hadron Collider (LHC) in July 2012 \cite{1,2}, understanding  the origin of Electro-Weak Symmetry Breaking (EWSB) is one of the primary tasks in particle physics. {EWSB in Beyond the Standard Models (BSM) scenarios can occur in two ways: through a weakly interacting model (with elementary Higgs bosons) or a strongly interacting one  (with composite Higgs bosons). } A composite Higgs state arising as a  pseudo Nambu-Goldstone Boson (pNGB), from a new dynamics at the compositeness scale $ f $, can lead to  EWSB{, as a result of  couplings of the Higgs sector  to the SM}. Therefore a  remarkable way to generate new light (pseudo)scalars in the spectrum is to make them also pNGBs.
We study here  a Composite 2-Higgs Doublet Model (C2HDM) \cite{3}, wherein two Higgs fields, $ \Phi_{1} $  and $ \Phi_{2}$, eventually inducing a SM-like Higgs state ($ h $), a CP-even field $(H)$, a CP-odd one $(A)$  and a charged pair $ (H^{\pm}) $, are introduced.

Perturbative unitarity bounds of a C2HDM based on $ SO(6)\rightarrow SO(4)\times SO(2) $ coset structure will be reported upon by using perturbative unitarity to cover all    energies  accessible at the LHC. The pNGB nature of the Higgses results in a modification of their couplings to matter with respect to the E2HDM (Elementary 2HDM) case and, consequently, the scattering amplitudes are left with a non-vanishing $s$-dependence. Here we will not present the Higgs potential generated by  radiative corrections. Instead,
we will assume for it the same structure as in the E2HDM, see our work in  Ref. \cite{DeCurtis:2016scv}. Moreover, in \cite{4}, we have studied LHC phenomenology of different  Yukawa types of a C2HDM. Herein, the theoretical and experimental  constraints on the parameter space  of   C2HDMs and   E2HDMs showed differences and the latter will guide us in the phenomenological study of a C2HDM versus a E2HDM. 

This report is organized as follows. In Section II  the C2HDM based on
$SO(6)/SO(4)\times SO(2)$ will be discussed. In section III, unitarity bounds will be discussed. In section IV,  
the LHC phenomenology  of   C2HDMs is presented while conclusions are given in section V.

\section{Effective Lagrangian in the C2HDM}
\subsection{ Kinetic Lagrangian and Gauge Couplings} 
In general, once the coset space has been chosen, the low-energy Lagrangian is fixed at the two-derivative level,  the basic ingredient being the pNGB matrix which transforms non-linearly under the global group.  

The kinetic Lagrangian invariant under the $SO(6)$ symmetry 
can be constructed by the analogue of the construction in non-linear sigma models
developed in \cite{ccwz}, as 
\begin{align}
\mathcal{L}_{\text{kin}}=\frac{f^2}{4}(d_\alpha^{\hat{a}})_\mu(d_\alpha^{\hat{a}})^\mu, ~~ \text{with} ~~ (d_\alpha^{\hat{a}})_\mu = i\,\text{tr}(U^\dagger D_\mu U T_\alpha^{\hat{a}} ),~~\text{where}~~ \alpha=1,2. ~~\hat{a}=1,4.  \label{ccwz2}
\end{align}
Here $U$ is the pNGB matrix:
\begin{align}
&U=\exp\left(i\frac{\Pi}{f}\right),~~\text{with}~~
\Pi\equiv\sqrt{2}h_{\alpha}^{\hat{a}}T^{\hat{a}}_{\alpha}=-i\left( \begin{array}{ccc} 0_{4\times 4} & h_1^{\hat{a}} & h_2^{\hat{a}}\\
-h_1^{\hat{a}} & 0  & 0 \\
-h_2^{\hat{a}} & 0 & 0 \end{array} \right). 
\label{u6}
\end{align}
In Eq.~(\ref{ccwz2}), the covariant derivative $D_\mu$ is given by 
\begin{align}
D_\mu &= \partial_\mu -ig T_L^a W_\mu^a -ig' YB_\mu, \quad \text{where} \quad T^{\hat{a}}, ~ T^{a}_{L}~ \text{and}~Y~\text{are}~SO(6)~\text{generators}. 
\end{align}
Some coefficients of the $ \Phi VV $ and $ \Phi \Phi V (V=W^{\pm} ,Z)$ couplings are given by
\begin{align}
&hW^{+}W^{-}=gm_{W}(1-\frac{\xi}{2})\cos \theta, \quad HW^{+}W^{-}=-gm_{W}(1-\frac{\xi}{2})\sin\theta,\\  &H^{+}\partial^{\mu}HW^{-}_{\mu}=\mp i\frac{g}{2}(1-\frac{5}{6}\xi)\cos\theta, \quad 
H^{+}\partial^{\mu}hW^{-}_{\mu}=\mp i\frac{g}{2}(1-\frac{5}{6}\xi)\sin\theta,\\
& A\partial^{\mu}hZ_{\mu}=- \frac{g_{z}}{2}(1-\frac{5}{6}\xi)\sin\theta,~\text{where}~\xi=\upsilon_{\rm  SM}^{2}/f^{2}~(\theta~\text{is the mixing angle between CP-even states)}.
\end{align}

\subsection{Yukawa Lagrangian}
The low-energy (below the scale $f$)  Yukawa Lagrangian is constructed
by determining  the embedding scheme of  the SM fermions into $SO(6)$ multiplets. 
This embedding can be explained via the mechanism based on the partial compositeness
hypothesis \cite{Kaplan:1991dc} by mixing  
elementary SM fermions  with composite fermions in the invariant form under the SM $SU(2)_L\times U(1)_Y$ gauge symmetry. The Yukawa Lagrangian is then given in terms of the {\bf 15}-plet of pNGB fields $\Sigma$ and the {\bf 6}-plet of fermions defined as \cite{4}
\begin{align}
{\cal L}_Y = f\Big[\bar{Q}_L^u (a_u \Sigma - b_u \Sigma^2) U_{R} 
+\bar{Q}_L^d (a_d \Sigma - b_d \Sigma^2) D_{R}
+\bar{L}_L (a_e \Sigma - b_e \Sigma^2) E_{R}\Big] + \text{h.c.} 
\end{align}
{We derive our Yukawa Lagrangian up to the squared power of $ \Sigma $, since the terms with the cubic and more than cubic power of $\Sigma$ do not generate any extra distinct contribution to it. After the Yukawa Lagrangian is described in terms of the complex doublet Higgs fields $ \Phi_{1} $ and $ \Phi_{2} $,  Flavor Changing Neutral Currents (FCNCs) occur 
in C2HDMs, as seen already in E2HDMs. In order to prevent FCNCs, we implement a $C_2$ symmetry} {transformation~\cite{3} as follows:
\begin{align}
&U(\pi_1^{\hat{a}},\pi_2^{\hat{a}}) \to C_2U(\pi_1^{\hat{a}},\pi_2^{\hat{a}})C_2 = U(\pi_1^{\hat{a}},-\pi_2^{\hat{a}}), 
\end{align}
where 
\begin{align}
C_2 = \text{diag}(1,1,1,1,1,-1).
\end{align} }
\begin{table}[h]
\begin{center}
\hspace*{-1.25truecm}
{\renewcommand\arraystretch{1}
\begin{tabular}{c||ccc||ccc||ccc|ccc|ccc}\hline\hline
 &$U_R$&$D_R$&$E_R$& $(a_u,b_u)$ & $(a_d,b_d)$ & $(a_e,b_e)$  & $\bar{X}_u^h$   & $\bar{X}_d^h$   & $\bar{X}_e^h$    & $\bar{X}_u^H$    
& $\bar{X}_d^H$    & $\bar{X}_e^H$   & $\bar{X}_u^A$& $\bar{X}_d^A$ & $\bar{X}_e^A$ \\\hline 
Type-I &$-$&$-$&$-$  & $(0,\surd)$ &$(0,\surd)$ & $(0,\surd)$ & $\zeta_h$ & $\zeta_h$ & $\zeta_h$  & $\zeta_H$  & $\zeta_H$ & $\zeta_H$ & $\zeta_A$ & $\zeta_A$ & $\zeta_A$  \\\hline 
Type-II&$-$&$+$&$+$  & $(0,\surd)$ &$(\surd,0)$ & $(\surd,0)$ & $\zeta_h$ & $\xi_h$   & $\xi_h$    & $\zeta_H$  & $\xi_H$   & $\xi_H$   & $\zeta_A$ & $\xi_A$   & $\xi_A$    \\\hline 
Type-X &$-$&$-$&$+$  & $(0,\surd)$ &$(0,\surd)$ & $(\surd,0)$ & $\zeta_h$ & $\zeta_h$ & $\xi_h$    & $\zeta_H$  & $\zeta_H$ & $\xi_H$   & $\zeta_A$ & $\zeta_A$ & $\xi_A$    \\\hline 
Type-Y &$-$&$+$&$-$  & $(0,\surd)$ &$(\surd,0)$ & $(0,\surd)$ & $\zeta_h$ & $\xi_h$   & $\zeta_h$  & $\zeta_H$  & $\xi_H$   & $\zeta_H$ & $\zeta_A$ & $\xi_A$   & $\zeta_A$  \\\hline \hline 
\end{tabular}}
\caption{Charge assignment for right-handed fermions under the $C_2$ symmetry in  C2HDMs. 
All the left-handed fermions $Q_L^u$, $Q_L^d$ and $L_L$ are transformed as even under $C_2$. }
\label{types}
\end{center}
\end{table}
and then interaction Lagrangian becomes:
\begin{align}
{\cal L}_Y 
&=\sum_{f=u,d,e}\frac{m_f}{v_{\text{SM}}}\bar{f}\Big(\bar{X}_f^h h +\bar{X}_f^H H - 2iI_f \bar{X}_f^A  \gamma_5 A \Big)f\notag\\
&+ \frac{\sqrt{2}}{v_{\text{SM}}}\bar{u}V_{ud}(m_d\bar{X}_d^A\,P_R-m_u\bar{X}_u^A\,P_L  )d\,H^+  + \frac{\sqrt{2}}{v_{\text{SM}}}\bar{\nu}m_e\bar{X}_eP_R\, e \, H^+ +\text{h.c.}  
\end{align} 
The definition of $\zeta_{h,H,A}$ and $\xi_{h,H,A}$ are given by (at the first order in $ \xi $): 
\begin{align}
\zeta_h & =  \left(1-\frac{3}{2}\xi \right)c_\theta + s_\theta \cot\beta, \quad \xi_h  =  \left(1-\frac{\xi}{2} \right)c_\theta  -s_\theta  \tan\beta, \label{xihh}\\
\zeta_H & =  -\left(1-\frac{3}{2}\xi \right)s_\theta + c_\theta \cot\beta, \quad \xi_H  =  -\left(1-\frac{\xi}{2} \right)s_\theta  - c_\theta  \tan\beta, \label{xibh} \\
\zeta_A & =  \left(1+\frac{\xi}{2} \right)\cot\beta, \quad 
\xi_A    =  -\left(1-\frac{\xi}{2} \right)\tan\beta. 
\end{align}
In the limit of $\xi\to 0$, 
these coefficients get the same form as the corresponding ones in a softly-broken (by a mass term $M$) $Z_2$ symmetric version of the E2HDM~\cite{typeX}. See Tab.~\ref{types} for the available Yukawa types of a 2HDM.
\section{ Unitarity Bounds}
\begin{figure}[h!]
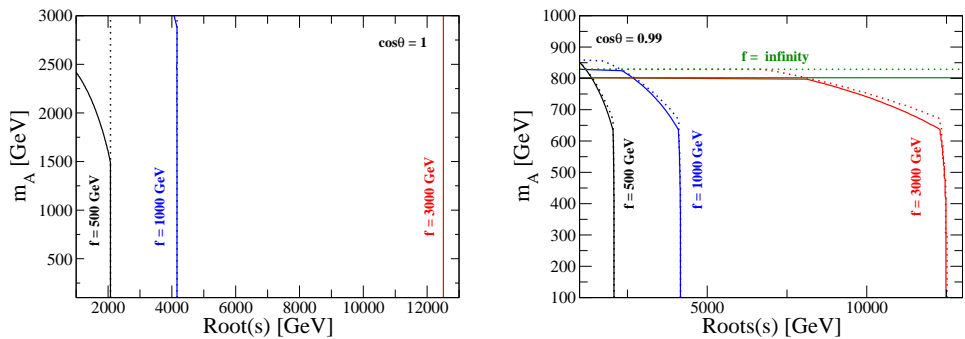

\begin{center}
\includegraphics[width=60mm]{bound_ct1.eps} \hspace{5mm}
\includegraphics[width=60mm]{bound_ct099.eps}
\caption{Constraints 
from unitarity and vacuum stability in the case of $\tan\beta = 1$ and $m_{H^\pm}^{} = m_A^{}$
for several fixed values of $f$.  
 $m_{H}^{}$ = $m_A^{}$ for the solid curves,  
 $m_{H}=m_A^{}\pm 500$ GeV for the dashed curves. 
For all the plots, $M$ is scanned. $ f\rightarrow \infty $ is E2HDM case.}
\label{bound_1}
\end{center}
\end{figure}

In order to calculate the bounds from perturbative unitarity in our C2HDMs, all  possible 2-to-2-body bosonic elastic scatterings are considered. Our conclusive results have been presented in \cite{DeCurtis:2016scv}. We use the same method as the  elementary models such as  the SM~\cite{LQT} and E2HDM~\cite{KKT,Akeroyd,Ginzburg_CPV,Ginzburg,KY} to find such unitarity bounds. {For example, we implement  the following condition }\cite{HHG} {for each eigenvalue of the $S$-wave amplitude matrix: 
\begin{align}
|\text{Re}(x_{i})|\leq 1/2.
\end{align}}
The most important difference between the unitarity bounds  in a E2HDM  and a C2HDM is the presence of a squared energy dependence in the $S$-wave amplitude of the latter, since in C2HDMs  the Higgs-Gauge-Gauge  couplings is modified from that in  E2HDMs by the factor $(1-\xi)$. The energy dependence of the $S$-wave amplitudes cause unitarity violation and asks for an Ultra-Violet (UV) completion of the C2HDM. However, we stay in the energy region where  C2HDMs and E2HDMs  are both unitarity in order to compare the bounds on the mass of the extra Higgs bosons. In Fig.\ref{bound_1}, one of the  main result is that  there is an upper limit on $ \sqrt{s} $ for the  no-mixing $ \cos\theta=1 $ case, while  there is an upper limit on both  $ \sqrt{s} $ and $ m_{\Phi} $ for non-zero mixing $ \cos\theta=0.99 $ case.\\

\section{ LHC phenomenology}
We first  study the constraints on the parameter space of C2HDMs from searches for  extra Higgs bosons at collider experiments and significant differences among the various   Yukawa types are reported for  the different values of   $ \xi=~ 0.1,~ 0.07,~0.04$ and 0 (which is E2HDM case) \cite{4}. We then investigate, over the accessible regions of parameter space,  the deviations of the SM-like Higgs ($h$) couplings from the SM values. The scaling factors $\kappa_X^{}$ for the $hXX$ couplings are given by $\kappa_X^{} = g_{hXX}^{\text{NP}}/g_{hXX}^{\text{SM}}$ and, in the C2HDM, these are given at the tree level by 
\begin{align}
\kappa_V^{} = \left(1-\frac{\xi}{2} \right)c_\theta~~(V=W,Z),\quad 
\kappa_f^{} = \bar{X}_f^h = \zeta_h\text{~~or~~}\xi_h~~(f=u,d,e). 
\end{align}   
In  C2HDMs  
non-zero values of both $\xi$ and $\theta$ can give $\kappa_X^{} \neq 1$, while in E2HDMs only that of $ \theta $ can give $\kappa_X^{}\neq 1$.  This can lead to deviations from the alignment limit obtained very differently in the two 
scenarios considered (e.g., it is seen that even for $ \theta=0 $, i.e., no mixing between the  CP-even  Higgs bosons $ h $ and $ H $, the $ hVV $ couplings of C2HDMs deviate from the SM values), yielding totally different phenomenology, both in production and decay of the heavy $H$, $A$ and $H^\pm$ states. For example, 
{ $\Delta\kappa_V^{}=-2\%$ is obtained by  $(\xi,\theta)=(0.04,0)$ and $(0,0.2)$ in a C2HDM and E2HDM, respectively.} As a result, we  can distinguish the two scenarios from the decay Branching Ratios (BRs) of the extra Higgs bosons
for a given value of $\Delta\kappa_V^{}=\kappa_{V}-1$. Therefore,  measurement of  a non-zero value of $\Delta\kappa_V^{}$  at the LHC would give an  indirect evidence for a non-minimal Higgs sector and, by focusing on the same value of $\Delta\kappa_V^{}$, we could find that  the dominant decay modes of the extra Higgs bosons $H$, $A$ and $H^\pm$ in C2HDMs are different from the  E2HDM ones. Similarly, the  production cross sections of in C2HDMs and E2HDMs, with the same value of $\Delta\kappa_V^{}$,  would show significant differences between the elementary and composite cases. Finally, deviations in $\Delta\kappa_f^{}$ could guide us in distinguishing amongst the possible types of Yukawa interactions. 

\begin{figure}[h!]
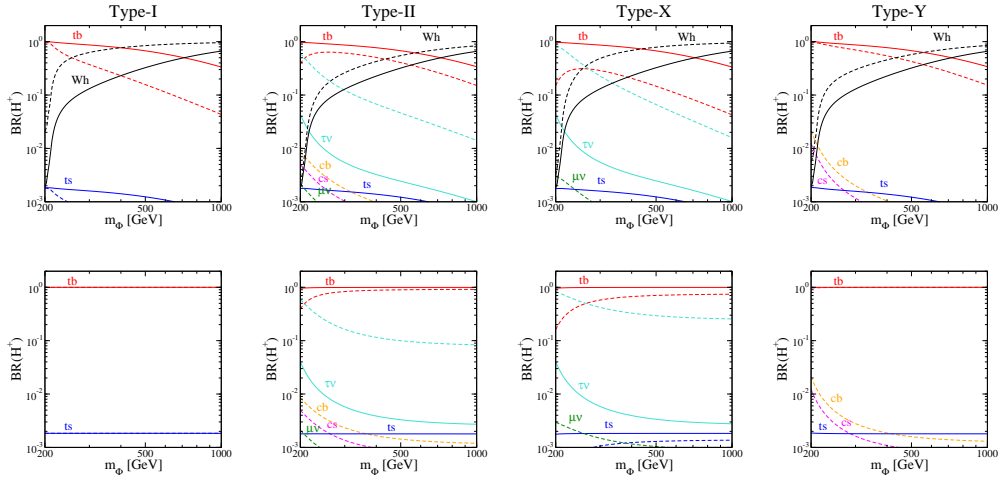

\begin{center}
\includegraphics[width=30mm]{BRm_Hp_1.eps}\hspace{3mm}
\includegraphics[width=30mm]{BRm_Hp_2.eps}\hspace{3mm}
\includegraphics[width=30mm]{BRm_Hp_3.eps}\hspace{3mm}
\includegraphics[width=30mm]{BRm_Hp_4.eps}\\ \vspace{5mm}

\includegraphics[width=30mm]{BRm_Hp_1_C.eps}\hspace{3mm}
\includegraphics[width=30mm]{BRm_Hp_2_C.eps}\hspace{3mm}
\includegraphics[width=30mm]{BRm_Hp_3_C.eps}\hspace{3mm}
\includegraphics[width=30mm]{BRm_Hp_4_C.eps}
\caption{BRs of $H^{\pm}$ as a function of $m_\Phi^{}(=m_H^{}=m_A^{}=m_{H^\pm})$ with $\tan\beta = 3(10)$ for solid (dashed) curves and $M = m_\Phi^{}$
in the four types of Yukawa interaction. 
The upper and lower panels show the case for $(s_\theta,\xi) = (-0.2,0)$ (E2HDM) and $ (s_\theta,\xi) = (0,0.04)$ (C2HDM), respectively. Here, $ \Delta\kappa_{V}=-2\% $.
}
\label{br_Hp2}
\end{center}
\end{figure}

As illustrative of this situation, we present Fig. \ref{br_Hp2}, wherein one can see that,  in the upper panels (referring to  the E2HDM), $ H^{\pm}\rightarrow W^{\pm}h $ or  $ H^{\pm}\rightarrow t\bar{b} $  are dominant for most cases, while, in the lower panels (referring to the C2HDM), only $H^{\pm}\rightarrow t\bar{b} $ is dominant. In fact, in the high mass region of the upper panels, $ H^{\pm}\rightarrow W^{\pm}h $ becomes the most dominant decay mode, while in the  lower panels $H^{\pm}\rightarrow t\bar{b} $ is the dominant mode in the whole mass region.

\section{Conclusions}
To conclude, we have discovered significant differences on the allowed parameter space of C2HDMs and E2HDMs from perturbative unitarity at the possible  energies of the LHC. In addition, possibilities of distinguishing the two sets of scenarios exist herein even if deviations from SM couplings of the SM-like Higgs state are identical in the two cases.

\section*{Acknowledgements}

SM and EY are supported in part through the NExT Institute.

\end{document}